\documentclass[superscriptaddress,aps,pra,twocolumn,showpacs,nofootinbib,longbibliography, floatfix]{revtex4-2}
\usepackage{braket}
\usepackage{amsfonts}
\usepackage{amsmath,amssymb,amsthm,amstext,mathrsfs}
\usepackage{amssymb}\usepackage[T1]{fontenc}
\usepackage{latexsym}
\usepackage{braket}
\usepackage{amssymb}
\usepackage[colorlinks=true,citecolor=blue,urlcolor=blue]{hyperref}
\usepackage{color}
\usepackage{graphics,epstopdf}
\usepackage{soul}
\usepackage[demo]{graphicx}
\usepackage{subcaption}
\usepackage{lipsum}
\usepackage{float}
\usepackage{adjustbox}
\usepackage[normalem]{ulem}
\usepackage{physics}
\usepackage{gensymb}

\newcommand{\be}{\begin{equation}}
\newcommand{\ee}{\end{equation}}
\newcommand{\ba}{\begin{eqnarray}}
\newcommand{\ea}{\end{eqnarray}}

\definecolor{ss}{RGB}{180,30,150}

\begin{document}
\title{Security of Device-independent Quantum Key Distribution under Sequential Attack}	

\author{Pritam Roy}
\email{roy.pritamphy@gmail.com}
\affiliation{S. N. Bose National Centre for Basic Sciences, Block JD, Sector III, Salt Lake, Kolkata 700 106, India}

\author{Souradeep Sasmal}
\email{souradeep.007@gmail.com}
\affiliation{Institute of Fundamental and Frontier Sciences, University of Electronic Science and Technology of China, Chengdu 611731, China}

\author{Subhankar Bera}
\email{berasanu007@gmail.com}
\affiliation{S. N. Bose National Centre for Basic Sciences, Block JD, Sector III, Salt Lake, Kolkata 700 106, India}
\affiliation{Department of Physics and Center for Quantum Frontiers of Research and Technology (QFort), National Cheng Kung University, Tainan 701, Taiwan}

\author{Shashank Gupta}
\email{shashankg687@gmail.com}
\affiliation{Okinawa Institute of Science and Technology Graduate University, Okinawa, Japan}
\affiliation{Indian Institute of Technology Indore, Khandwa Road, Indore, Simrol 453552, India}

\author{Arup Roy}
\email{arup145.roy@gmail.com}
\affiliation{Department of Physics, Hooghly Mohsin College, Chinsurah, West Bengal 712101, India}

\author{A. S. Majumdar}
\email{archan@bose.res.in}
\affiliation{S. N. Bose National Centre for Basic Sciences, Block JD, Sector III, Salt Lake, Kolkata 700 106, India}

\begin{abstract}
Device-independent quantum key distribution (DI-QKD) leverages nonlocal correlations to establish cryptographic keys between two honest parties while making minimal assumptions about the underlying systems. The security of DI-QKD is grounded in the validity of quantum theory, with Bell violations ensuring the intrinsic unpredictability of observed statistics, independent of the trustworthiness of the devices. While traditional collective QKD attacks assume that the adversary prepares the shared system, we analyse a scenario where the adversary does not control the source and instead interacts sequentially with the travelling system. In this setting, Eve performs an unsharp measurement that produces effective noise while preserving the observed Bell violation. Although such behaviour is already accounted for in existing DI-QKD security proofs, examining it through an explicit sequential interaction offers a concrete and physically motivated example of how these effective statistics can arise in practice. Our analysis further shows that, within a specific parameter regime, this sequential strategy reproduces some features of an optimal collective attack.
\end{abstract}

\maketitle


\section{Introduction} 

The aim of key distribution is to securely share a secret key between two parties, say Alice and Bob, to enable confidential communication. In accordance with Kirkhoff's Principle-`the enemy knows the system'-the security of any cryptosystem must rely exclusively on the secrecy of the key, rather than on the algorithm, method, or device employed \cite{Shannon1949, Renner2022}. Beyond the foundational insights into the interpretation of quantum theory, the violation of Bell inequalities offers the requisite device independence for key distribution protocols, ensuring security even when the devices utilised are untrusted \cite{Acin2006njp, Acin2006, Ekert2014, Brunner2014}. 

Bell's work \cite{Bell1964, Bell1966} elucidates that when quantum statistics violate a Bell inequality, it implies that such statistics cannot be generated by any pre-shared strategy—more precisely, through Local Operations and Shared Randomness (LOSR) \cite{Buscemi2012,Schmid2023}. This renders the statistics inherently unpredictable, irrespective of the adversary's knowledge of the underlying processes  \cite{Cavalcanti2012}. Since violations of Bell inequalities can be verified solely through observed statistics, this verification serves as a black-box test of unpredictability, thereby pioneering the physics behind Device-Independent Quantum Key Distribution (DI-QKD) with minimal assumptions \cite{Devetak2005, Acin2006njp, Acin2006, Acin2007, Vazirani14, Friedman2018, Pirandola2020, Sekatski2021, Schwonnek2021, Zapatero2023, ghoreishi2025futureDIQKD, Wooltorton2024, Farkas2024}. The security of DI-QKD is thus predicated solely on the validation of quantum theory, allowing users to forgo trust in the inner workings of the devices and ensuring security against any adversary constrained by quantum theory.

Traditional QKD attacks typically focus on the quantum channel during transmission and can be categorised into individual \cite{Ekert1994, Slutsky1998, Ltkenhaus1999}, collective \cite{Acin2007, Pironio2009, Roy2026Collective_QKD}, and coherent \cite{Shor2000} attacks. In these models, Eve is allowed to distribute the shared state or hold its purification. One might contend that if one party prepares the quantum system and shares a subsystem with the other, Eve, the adversary, would lose control over the preparation device, thereby diminishing her ability to predict the generated key. Additionally, with only one particle traversing the channel instead of two, the noise could be reduced, potentially enhancing the noise tolerance of DI-QKD protocols.

However, by manipulating the quantum channel, Eve could regain control. This motivates the question of what explicit sequential operations on the travelling system are compatible with the statistics expected in DI-QKD. In this work, we analyse an attack where Eve applies an unsharp measurement. While this does not allow her to gain any more information at a given noise level than what would be covered by existing security proofs, it may be physically easier to implement than the optimal collective attack. This provides a concrete example of an operation Eve can perform, though it does not affect the protocol's security.
 
Intuitively, one might expect that Eve’s measurement would disturb the quantum state, preventing Alice and Bob from observing a violation of a Bell inequality, such as the CHSH inequality \cite{Clauser1969} in the simplest scenario involving two parties, two measurements per party, two outcomes per measurement (2-2-2). On the contrary, recent findings reveal that Bell nonlocality can indeed be shared between Alice and multiple independent Bobs through sequential violations of the CHSH inequality based on unsharp measurements \cite{Silva2015, Mal2016, Brown2020, Steffinlongo2022, Sasmal2024, Cai2024}. 

Recently, there has been a notable surge of interest in the sequential sharing of quantum correlations, especially in areas such as entanglement \cite{Foletto2020, Pandit2022, ADas2022, Hu2023}, steering \cite{Sasmal2018, Shenoy2019, Han2024, Zhu2022, Rong2024}, multi-setting Bell nonlocality \cite{Das2019}, network nonlocality \cite{Mahato2022, Wang2022, Mao2023, Guo2024}, coherence \cite{Datta2019}, and contextuality \cite{Kumari2019, Kumari2023}, with extensions to multipartite scenarios  \cite{Saha2019, Maity2020, ShashankEPR2021,Xi2023, Zhang2024}. Several experiments have corroborated this phenomenon \cite{Schiavon2017,Hu2018, Feng2020, Foletto2020, Xiao2024, Virzi2024}, underscoring its significance to both foundational quantum research and practical applications. Applications of sequential quantum correlations encompass randomness generation \cite{Curchod2017, Foletto2021, Liu2024, Padovan2024}, reducing classical communication costs \cite{Tavakoli2018}, quantum teleportation \cite{Roy2021}, random access codes \cite{Mohan2019, Das2021}, self-testing of unsharpness parameters \cite{Anwer2021, Mukherjee2021, Roy2023, Paul2024, Roy2026}, and Remote state preparation \cite{Datta2024}. Existing studies establish that sequential quantum measurements can be implemented reliably in realistic setups. These results motivate us to present a specific sequential measurement-based attack relevant for DI-QKD that aligns with established security proofs.

Here, we consider an attack in which Eve applies an unsharp measurement. While this does not allow her to gain any additional information at a given noise level beyond what is already covered by existing security proofs, it provides a physically transparent and operationally explicit realization of how such effective statistics may arise in practice, in a manner analogous to explicit measurement-based attack models previously studied in prepare-and-measure QKD protocols \cite{Curty2005, Bechmann2006, Scarani2009, Renner2022}. Our analysis involves estimating the key rate with suitable quantifiers \cite{Devetak2005, Acin2007, Konig2009}, sequential CHSH violations, and the quantity of Quantum Bit Error Rate (QBER). We investigate two pertinent scenarios under one-way communication: (1) a sequential unsharp measurement attack where Alice sends the entangled particle to Bob, and Eve intervenes prior to its arrival at Bob's Lab (Sec.~\ref{SeqA}), and (2) the cumulative effect of the sequential and collective attacks (Sec.~\ref{sacacom}). 

In a standard collective attack, Eve controls the source, distributing a mixed entangled state to Alice and Bob while retaining its purification in a quantum memory for subsequent collective measurement. In a sequential attack, Eve intercepts the photon sent from Alice to Bob in each round, performs a measurement (Sec.~\ref{SeqA}) to maximise information gain while preserving the non-locality between Alice and Bob, and forwards the post-measurement state to Bob. In our analysis, this is used solely as a concrete sequential CPTP map compatible with the observed Bell violation, and it does not extend Eve’s capabilities beyond the standard DI-QKD adversarial model. We extend this approach to analyse the impact of sequential attack in the presence of a quantum memory on the key rate (Sec.~\ref{sacacom}). In this scenario, Eve performs the sequential measurement, prepares the ancilla states based on the observed outcome statistics, and stores the ancilla for a subsequent collective measurement, thereby integrating the characteristics of a sequential attack strategy. This hybrid model remains fully covered by established DI-QKD security proofs, and our role is to give a clear, physically grounded realisation of how such behaviour can occur.

To lay the foundations of our findings, we first validate the sequential attack and evaluate the secure key rate under the assumption that Eve employs unsharp measurements with a biased parameter to maximize information gain. This step merely specifies one of the physically realisable sequential strategies and does not alter the adversarial model assumed in standard DI-QKD security proofs. Our results show that, when the sequential attack is combined with collective attacks, Eve's knowledge becomes a non-linear function of the biased parameter. In particular, there exists a parameter region in which the resulting key rate coincides with that of a standard collective attack, even though Eve does not control the preparation device; this agreement is expected, as collective attacks already describe the optimal bound. To demonstrate this, we begin with a brief recap of the collective attack framework \cite{Acin2007, Pironio2009} (Sec.~\ref{ColA}), which serves as a reference for the analysis presented in the subsequent sections.


\section{DI-QKD under collective attacks}\label{ColA} 

Alice prepares a bipartite entangled state, $\rho \in \mathscr{L}(\mathcal{H}_A \otimes \mathcal{H}_B)$, and shares one-half with Bob via a quantum channel. This process spans several rounds $N$, each assumed independent, resulting in a shared state $\otimes_N \rho$. In each round, Alice randomly selects one of $m_A$ local measurements, $\qty{A_{x} : x\in [m_A]}$, recording outcomes $a \in \{1,2,..., d_A\}$. Similarly, Bob selects from $m_B$ local measurements, $\qty{B_y : y \in [m_B]}$, with outcomes $b \in \{1,2,..., d_B\}$. The resulting statistics are captured by a vector in real number space, $\mathscr{P}\equiv \{p(a,b|x,y)\} \in \mathbb{R}^{m_Am_Bd_Ad_B}$, where $P(a,b|x,y)$ denotes the joint conditional probability of obtaining outcome-pair $(a,b)$. 

In quantum theory, measurements $A_x$ and $B_y$ are described by Positive-Operator-Valued-Measures (POVMs), $A_x\in \mathscr{L}(\mathcal{H}^A)\equiv\{A_{a|x}\}$ and $B_y\in \mathscr{L}(\mathcal{H}^B)\equiv\{B_{b|y}\}$, respectively. The joint conditional probability follows the Born rule: $p(a,b|x,y,\rho) = \text{Tr}[\rho A_{a|x}\otimes B_{b|y}]$, where $A_{a|x}$ and $B_{b|y}$ are corresponding projectors for outcomes $a$ and $b$. After measurement, Alice and Bob publicly disclose their measurement settings and a subset of outcomes to estimate joint statistics, crucial for detecting violations of the Bell inequality. Such violations of a Bell inequality underpin the establishment of a private key between Alice and Bob \cite{Ekert1991}. 

Here, we consider that Alice selects from three measurement settings $(A_0, A_1, A_2)$ and Bob from two $(B_1, B_2)$. Using $A_1, A_2$ and $B_1, B_2$ respectively, Alice and Bob test nonlocality in the simplest 2-2-2 Bell scenario by observing violations of the following CHSH inequality \cite{Clauser1969}
\begin{equation}\label{S}
\mathcal{S} = \langle A_1 B_1 \rangle + \langle A_1 B_2 \rangle + \langle A_2 B_1 \rangle - \langle A_2 B_2 \rangle \underset{L}{\leq} 2
\end{equation}
where $\langle A_x B_y \rangle = \sum_{a,b} (-1)^{a+b} p(a,b|x,y,\rho)$ with $a,b\in\{+1,-1\}$.

Finally, after discarding the dataset used in the spot-checking phase, Alice and Bob extract a raw key from the pair $(A_0,B_1)$. In the presence of eavesdropping, the information accessible to Eve is bounded by the parameters $\mathcal{S}$ and the quantum bit error rate (QBER), $Q=p(a\neq b|A_0,B_1,\rho)$ \cite{Acin2007}. Subsequently, Alice and Bob distil a secure key from the raw data through classical post-processing via a public channel \cite{Acin2007}. 

In the context of collective attacks, Eve holds a purification of the shared state and interacts with each signal independently and identically. In this setting, when only considering one-way protocols, the Devetak–Winter formula~\cite{Devetak2005} provides the asymptotic secret key rate,
\begin{equation} \label{key}
r \geq 1 - H(Q) - \chi(B_1 : E),
\end{equation}
Here, $H$ is the binary entropy with $H(x) = -x\log_2 x - (1-x)\log_2 (1-x)$, and $H(Q) = 1 - \mathcal{I}(A_0 : B_1)$  where $\mathcal{I}(A_0 : B_1)$ is the mutual information between Alice and Bob. $\chi(B_1 : E)$ denotes the Holevo quantity, given by $\chi(B_1 : E)= S(\rho_E) - \frac{1}{2}\sum_{b_1 = \pm 1}S(\rho_{E|b_1})$, between Eve and Bob, which is upper bounded as $\chi(B_1 : E) \leq H\left(\frac{1+\sqrt{(\mathcal{S}/2)^2 - 1}}{2}\right)$, with $\mathcal{S}$ denoting the CHSH value \cite{Acin2007}. 

Consider a scenario where Alice prepares a maximally entangled two-qubit state, $\rho=\ket{\Phi^{+}}\bra{\Phi^{+}}$ and shares one-half of the entangled qubit with Bob through a quantum channel. They intend to implement the following optimal measurement directions to achieve the optimal CHSH value $2\sqrt{2}$.
\begin{equation}\label{MeasColl}
A_0 = B_1= \sigma_z ;  \ A_x = \frac{1}{\sqrt{2}}\qty[B_1+(-1)^xB_2] ; \ B_2 = \sigma_x ;
\end{equation}
where $x\in\{1,2\}$.  The measurement pair $(A_0,B_1)$ is employed by Alice and Bob for the evaluation of the QBER.  In the presence of a collective attack by Eve, the prepared state  is assumed to take
the form of a mixed entangled two-qubit state, given by
\begin{eqnarray}
\tilde{\rho} &=& \frac{1}{4}\qty(2+\sqrt{1-\alpha^2})\ket{\Phi^{+}}\bra{\Phi^{+}} + \frac{1}{4}\ket{\Phi^{-}}\bra{\Phi^{-}} \nonumber \\
&&+ \frac{1}{4}\qty(1-\sqrt{1-\alpha^2})\ket{\Psi^{+}}\bra{\Psi^{+}} \ \ \forall \alpha\in[0,1]\label{state1}
\end{eqnarray}
Note that in \cite{Acin2007}, a specific example of a two-qubit Werner state was used to compute the key rate. Similarly, in our work, we have selected a different class of two-qubit mixed entangled states, described by Eq.~(\ref{state1}), to evaluate the key rate. In what follows, we demonstrate that a positive key rate can also be achieved with this class of mixed states, despite a small CHSH violation. We focus on this particular class of mixed states to compare the key rate obtained under the collective attack scenario with that derived in the sequential attack model, which we introduce in the subsequent section.

\begin{figure}[h!]
\includegraphics[width=\linewidth]{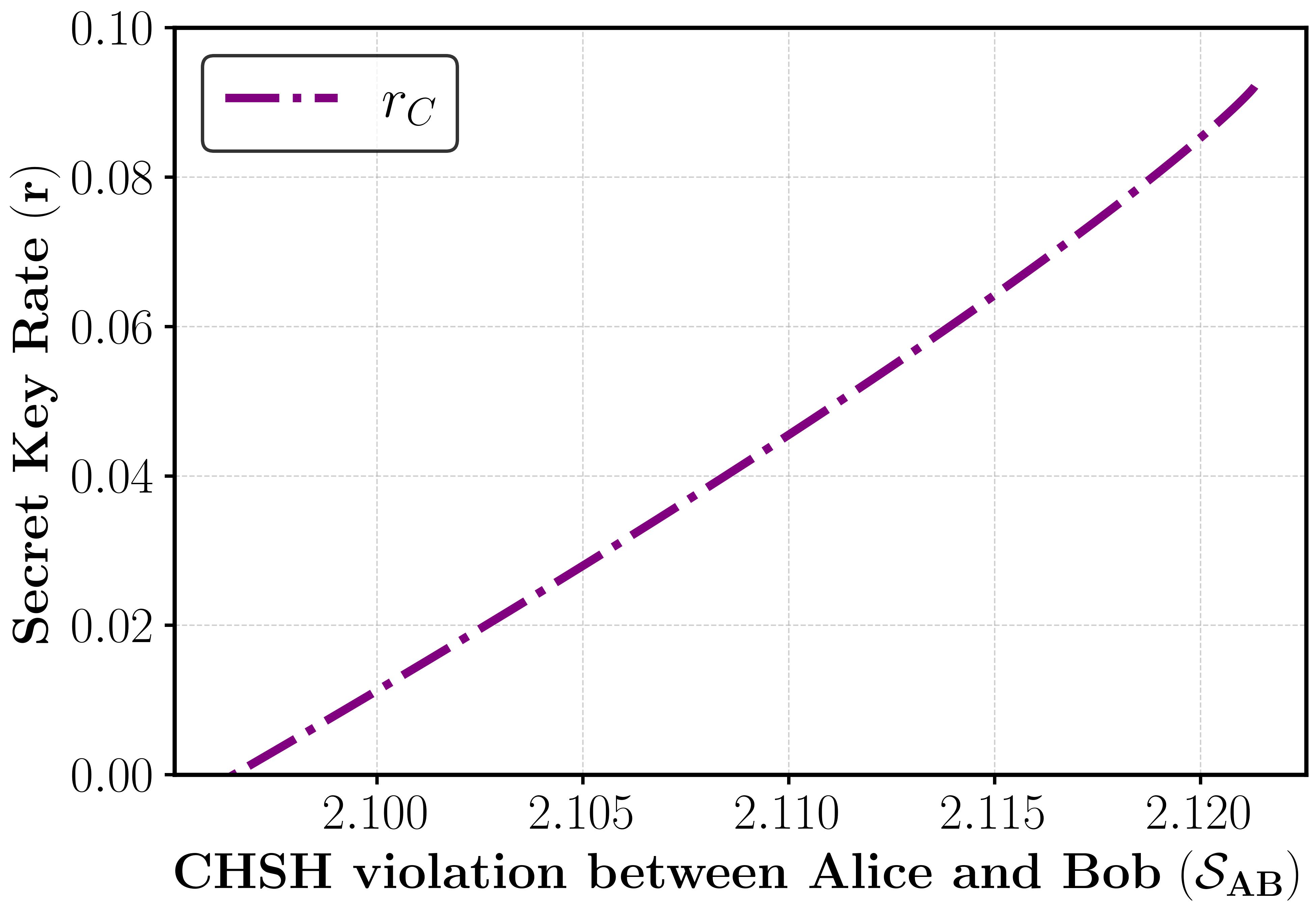}
\caption{Key rate ($r$) as a function of the CHSH violation between Alice and Bob ($S_{AB}$). The purple dashed-dotted is the key rate under a collective attack.}
\label{collrvsS}
\end{figure}

In this scenario, the CHSH value between Alice and Bob is given by
\begin{equation}
\mathcal{S}_{AB} = \frac{1}{\sqrt{2}}\left(2 + \sqrt{1-\alpha^2}\right)
\label{BellABColl}
\end{equation}
For ensuring CHSH violation, the parameter $\alpha$ is bounded by $\alpha\in[0,\sqrt{8 \sqrt{2}-11}\approx 0.5601)$. In this range, the CHSH value is constrained to the interval $\mathcal{S}_{AB}\in (2, \frac{3}{\sqrt{2}}\approx 2.1213]$. The QBER is evaluated as
\begin{equation} \label{qber}
Q = p(a\neq b|A_0,B_1,\tilde{\rho})=\frac{1}{4}\qty(1-\sqrt{1-\alpha^{2}})
\end{equation}
The key rate is evaluated from Eq.~(\ref{key}) by employing Eqs.~(\ref{BellABColl}) and (\ref{qber}). The lower bound of the key rate, denoted by $(r_{C})$, represents the minimum security threshold in the worst possible case. It is evaluated by taking the upper bound of $\chi(B_1:E)$, expressed as a function of QBER and the CHSH value, as given by \cite{Acin2007}
\begin{equation}
    r_{C} (Q,\mathcal{S}) = 1 - H(Q) - H\qty(\frac{1}{2}\qty[1+\sqrt{(\mathcal{S}_{AB}/2)^2 - 1}])
    \label{KeyRColl}
\end{equation}
Note that $r_{C} (Q,\mathcal{S})$ is found to be a monotonic increasing function of the CHSH value, thereby establishing that an increase in nonlocality will result in an increasing lower bound of the key rate as illustrated by curve `b' of Fig.~\ref{collrvsS}. The lower bound of the key rate is non-zero, i.e., $r_{C} (Q,\mathcal{S})\in (0,0.0921]$ for $Q\in[0,0.0086)$ and $\mathcal{S}_{AB}\in[2.0965,2.1213]$, implying $\alpha\in (0,0.2625]$.

In the following analysis, we focus on a concrete experimentally motivated sequential intervention on the travelling qubit, where Eve performs an unsharp measurement without controlling the preparation device, as a physically explicit example within the standard DI-QKD security framework.


\section{DI-QKD under Sequential attack}\label{SeqA}

Let us first describe the sequential measurement used in our analysis and how it affects the information available to Eve.

In this setting, Eve intercepts the qubit sent by Alice to Bob after the preparation of the maximally entangled state 
$\rho = \ket{\phi^{+}}\bra{\phi^{+}}$ during the transmission phase. 
To remain hidden, she performs an unsharp measurement chosen so that the CHSH violation observed by Alice and Bob matches that of Eq.~(\ref{BellABColl}). 
Although one might expect such an intervention to destroy the violation, a series of recent results has shown that Bell nonlocality can be sequentially shared among multiple observers through suitably tuned unsharp measurements \cite{Silva2015, Mal2016, Brown2020, Sasmal2024, Cai2024}. In our analysis, this serves only to specify a sequential measurement strategy consistent with the observed Bell violation and QBER, without affecting the security of the DI-QKD protocol. A schematic representation of this sequential interaction is provided in Fig.~\ref{ExptSetUp}.

\subsection{Proposed setup for the sequential attack strategy}\label{ExptProp}

\begin{figure*} 
\centering\includegraphics[width=\textwidth]{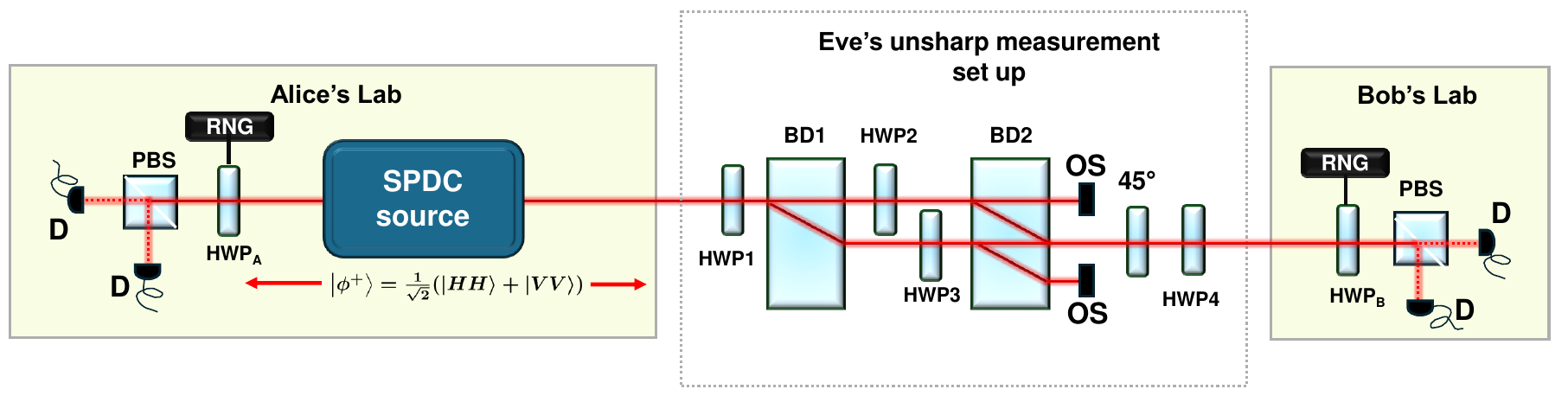}
    \caption{Proposed setup for DI-QKD under sequential attack. Entangled photon pairs are generated via SPDC and sent to Alice and Bob. Eve performs unsharp measurements using a combination of Beam Displacers (BD), Half-Wave Plates (HWP) to simulate weak and strong measurements. Alice and Bob randomly select measurement settings, rotating their HWPs to measure in different bases, followed by detection through Polarising Beam Splitters (PBS) and detectors (D).}
\label{ExptSetUp}
\end{figure*}

The sequential attack can be realized using the setups employed in recent experimental demonstrations of the sequential sharing of nonlocal correlations \cite{Hu2018, Schiavon2017, Xiao2024ExptSharing}. This attack can be implemented with a straightforward optical experimental setup, as depicted in Fig.~\ref{ExptSetUp}. The setup comprises the following four distinct stages:

(i) \textit{Quantum state preparation.--} Entangled photon pairs are generated via spontaneous parametric down-conversion (SPDC) in nonlinear crystals, such as periodically poled lithium niobate (PPLN) or beta barium borate (BBO) crystals \cite{kwiat1995, ursin2007}. Alice retains one photon from each pair and sends the other photon to Bob’s laboratory.
    
(ii) \textit{Eve's measurement set up.--} Eve employs von Neumann-type measurement frameworks \cite{Busch2016, Silva2015} to perform unsharp measurements on the photon directed to Bob's lab. For an unsharp measurement of the projector in the direction $\beta$, $\sigma_{\beta} = \ket{f(\beta)}\bra{f(\beta)} - \ket{f(\beta_\perp)}\bra{f(\beta_\perp)}$, Eve performs an optical transformation. Measuring $(\sigma_{\beta})$ involves rotating two half-wave plates (HWPs), (\text{HWP}{1}) and (\text{HWP}{4}), by $(\frac{\beta}{2})$, where $f(\beta) = \cos(\beta)\ket{H} + \sin(\beta)\ket{V}$. Here, $\ket{H}$ and $\ket{V}$ represent the horizontal and vertical polarization states of light, respectively. A Beam Displacer (BD) establishes weak coupling between the system and the pointer. The angles for (\text{HWP}{2}) and (\text{HWP}{3}) are set to $(\frac{\phi}{2})$ and $(\frac{\pi}{4} - \frac{\phi}{2})$, respectively. The quality factor $F=\sin{2\phi}$ enables $(\phi)$ to vary from $(\frac{\pi}{4})$ to $0$, transitioning from weak to strong coupling. Outcomes of $+1$ or $-1$ are determined by rotating (\text{HWP}{1}) and (\text{HWP}{4}) by $(\frac{\beta}{2})$ or $(\frac{\beta}{2} + \frac{\pi}{4})$, respectively, and, positioning optical shutters (OS) and beam displacers appropriately \cite{Hu2018}. By this method, Eve performs the desired measurements: $(\sigma_z)$ with bias $q$ by setting $(\phi = 0)$ (strong measurement, $\gamma_1 = 1)$, and $(\gamma\sigma_x)$ with bias $1-q$ by choosing $(\phi = \phi_0)$ (weak measurement, $\gamma_2 = \gamma = \cos 2\phi)$.

(iii) \textit{Alice's measurement.--} Alice selects her measurement basis randomly  using a quantum random number generator and chooses one of the three measurements specified in Eq.~(\ref{modam}). She then rotates her half-wave plate $(\text{HWP}_{A})$ by angles of $0$, $\frac{\theta}{2}$ or $\frac{\pi}{2} - \frac{\theta}{2}$, corresponding to $A_0, A_1$ and $A_2$, respectively, where $\theta$ is predetermined by Eve. 

(iv) \textit{Bob's measurement.--} Bob also selects a measurement basis randomly and chooses one of the two measurements specified in Eq.~(\ref{MeasColl}). He then rotates his half-wave plate $(\text{HWP}_{B})$ by an angle of $0$ or $\frac{\pi}{4}$, corresponding to measurements of $\sigma_z$ and $\sigma_x$, respectively.

After the photons pass through $\text{HWP}_{A}$ or $\text{HWP}_{B}$, Alice and Bob perform strong measurements on their respective photons using a Polarising Beam Splitter (PBS) coupled with a detector. 

It is to be noted that a quantum adversary’s setup possesses all the capabilities of Alice and Bob's setups and may be further augmented by quantum computational capabilities. Such enhancement would enable the implementation of advanced and optimised versions of such attacks (see Fig.~\ref{ExptSetUp}) in future scenarios.


\subsection{Sequential attack strategy} \label{sas}

It may be emphasised that since Alice is using $A_0$ to generate the raw key with Bob, the observable $E_1$, defined here as Eve’s measurement aligned with $A_0$, is the setting from which she would ideally extract the most information. However, performing the sharp measurement $E_1=A_0$ in every round would significantly disturb the shared state and remove the CHSH violation, causing the protocol to abort. Hence, $E_1$ is informationally favourable for Eve only when combined with the constraint that the Bell violation observed by Alice and Bob must remain above the threshold.

To satisfy both conditions, Eve instead performs an unsharp measurement corresponding to $E_1$ with probability $q$ and another unsharp observable $E_2$ with probability $1-q$, where $q>1-q$. The role of $E_2$ is to mitigate the disturbance caused by $E_1$ so that the overall statistics preserve the target CHSH value. To maximise the information compatible with this requirement, Eve chooses $q$ as large as allowed, adjusts Alice’s remaining measurement directions $(A_1, A_2)$ by an angle $\theta$, and selects unsharpness parameters $(\gamma_1, \gamma_2)$ such that the CHSH violation between Alice and Bob remains unchanged, as given in Eq.~(\ref{BellABColl}). The unsharp measurement performed by Eve is characterised by \cite{busch1996}.
\begin{equation}
    M_e \equiv\qty{\mathcal{E}_{g|e}=\frac{1}{2}\qty(\openone+\gamma_e E_e); \ e\in\{1,2\}} 
    \label{EveM}
\end{equation}
where $E_1=A_0=\sigma_z$ and $E_2=\sigma_x$; $g\in\{0,1\}$. Alice's rotated measurement directions $(A_1,A_2)$ are given by
\begin{equation}
A_1 = \cos{\theta}\sigma_{z}+\sin{\theta}\sigma_{x} \ ; \ A_2 = \cos{\theta}\sigma_{z}-\sin{\theta}\sigma_{x}
\label{modam}
\end{equation}

Next, Eve transfers the post-measurement qubit to Bob. During the device-independence test, when Alice publishes her outcome, Eve collects Alice's outcome statistics and evaluates the joint probability distribution as $p(a,g|A_x,M_e)=\Tr[\rho A_{a|x} \otimes \mathcal{E}_{g|e}]$. As a result of Eve's unsharp measurement in a biased input manner, the state evolves into a mixed entangled two-qubit state, $\rho_1$, as described by Luder's transformation rule \cite{busch1996}. Since Eve performs two distinct measurements ($M_{e}$) on $\rho$, the post-measurement states, denoted by $(\rho)_{M_e}$, produced with probabilities $q$ and $1-q$, and are given by
\begin{equation} \label{evem}
(\rho)_{M_e} \rightarrow{} \sum_{g} \qty(\mathbb{I} \otimes \sqrt{\mathcal{E}_{g|e}} ) \rho \qty(\mathbb{I} \otimes \sqrt{\mathcal{E}_{g|e}}) 
\end{equation}
Since Bob doesn't have any information about Eve's unsharp measurement, the post-measurement state $\rho_1$ after Eve's measurement is 
\begin{equation} \label{indbob}
\rho_1= q \ (\rho)_{M_1} + (1-q) \ (\rho)_{M_2}
\end{equation}
To evaluate the quantity $\sqrt{\mathcal{E}_{g|e}}$ in Eq.~(\ref{evem}), we take recourse to the Kraus-operator formalism \cite{busch1996}. The Kraus operators are given by 
\begin{equation}
\mathbf{K}_e \equiv \qty{\mathcal{K}_{g|e} \Big| \mathcal{K}_{g|e}^{\dagger}\mathcal{K}_{g|e}\geq0, \ \sum_g \mathcal{K}_{g|e}^{\dagger} \mathcal{K}_{g|e} =\openone} 
\end{equation}
We construct Kraus operator such a way, so that $\qty(\mathcal{K}_{g|e})^{\dagger} \qty(\mathcal{K}_{g|e})=\mathcal{E}_{g|e}$ and, in turn, $\sqrt{\mathcal{E}_{g|e}}=\mathcal{K}_{g|e}$. The Kraus operator in this scenario is then given as follows
\begin{equation}
\mathcal{K}_{g|e}= K_0 \openone  +(-1)^g K_1 E_e 
\end{equation}
where $K_i=\frac{\sqrt{1+\gamma_e}+(-1)^i\sqrt{1-\gamma_e}}{2\sqrt{2}}$ with $i \in \{0,1\}$. The average reduced state after Eve's unsharp measurement given by Eq.~(\ref{indbob}) is evaluated as
\begin{eqnarray}\label{state2}
\rho_{1}&=&  \sum_{e}\sum_{g}q_e\qty(\openone \otimes \mathcal{K}_{g|e}) \rho \qty(\openone \otimes \mathcal{K}_{g|e}) \nonumber \\
&= & \frac{1}{2}\qty[1+q\sqrt{1-\gamma_{1}^{2}}+(1-q)\sqrt{1-\gamma_{2}^{2}}]\ket{\Phi^{+}}\bra{\Phi^{+}} \nonumber \\ 
&& + \frac{q}{2}\qty(1-\sqrt{1-\gamma_{1}^{2}})\ket{\Phi^{-}}\bra{\Phi^{-}} \nonumber \\ 
&& + \frac{1-q}{2}\qty(1-\sqrt{1-\gamma_{2}^{2}})\ket{\Psi^{+}}\bra{\Psi^{+}}
\end{eqnarray}
Note that with these choices of measurements by Eve and Alice, the nonlocality between them will be ensured by the violation of a biased Bell inequality. The biased Bell inequality \cite{Lawson2010} where one party measures in an unbiased manner and the other party measures with an input bias $q\geq \frac{1}{2}$ is expressed as 
\begin{eqnarray} 
\mathcal{S}_{Biased}  &=& \frac{q}{2}\qty[\langle A_0 B_0\rangle - \langle A_0 B_1\rangle + \langle A_1 B_0\rangle + \langle A_1 B_1\rangle] \nonumber \\
     &&+ \frac{1}{2}\qty(\langle A_0 B_1\rangle + \langle A_1 B_1\rangle) - \langle A_1 B_1\rangle\underset{L}{\leq} q
     \end{eqnarray} 
The corresponding nonlocality between Alice and Eve is given by     
\begin{equation}
     \mathcal{S}_{AE} =\gamma_1 q  \cos{\theta}  + \gamma_{2} (1 - q) \sin{\theta} 
\end{equation}
Nonlocality between Alice and Eve is ensured for the following conditions
\begin{equation} 
\begin{cases}
\gamma_1\cos\theta-\gamma_2 \sin\theta>1 \ \text{if} \ \frac{1}{2}\leq q \leq 1 \\
\gamma_1\cos\theta-\gamma_2 \sin\theta<1 \ \text{if} \   \frac{1}{2} \leq q<\frac{\gamma_2\sin\theta}{\gamma_1\cos\theta-\gamma_2 \sin\theta-1} \nonumber
\end{cases}
\end{equation}

Note that the parameter $q$ may take overlapping values under different conditions when nonlocality is analysed independently of $\gamma_1$, $\gamma_2$ and $\theta$. However, in our scenario, ensuring Bell violation between Alice and Eve requires a trade-off among these parameters, as described by the above equation. Additionally, further constraints must be imposed to ensure CHSH violation between Alice and Bob.

For the reduced state $\rho_1$, the CHSH value between Alice and Bob is
\begin{equation}
\begin{aligned}
    \tilde{\mathcal{S}}_{AB} & = 2 \Bigg[ \sin{\theta} - q \sin{\theta} \qty(1 - \sqrt{1 - \gamma_{1}^{2}} )  \\ 
    &  + \cos{\theta} \qty( q + \sqrt{1 - \gamma_{2}^{2}} - q \sqrt{1 - \gamma_{2}^{2}} ) \Bigg]
\end{aligned}
\end{equation}

It needs to be mentioned that even if Eve chooses her first measurement, $E_1=A_0$, to maximise the information gain, the CHSH value between Alice and Bob persists. In this case, for $\gamma_1=1$, the CHSH value between Alice and Bob reduces to the following expression (writing $\gamma_2=\gamma$ from now on)
\begin{equation}\label{BellABSeq}
    \tilde{\mathcal{S}}_{AB} = 2 \bigg[ q \cos\theta + \qty(1 - q )\qty(\sin{\theta}+\cos\theta \sqrt{1 - \gamma^{2}}) \bigg]
\end{equation}
Both Bob and Eve demonstrate simultaneous nonlocality with Alice under the following conditions
\begin{equation} \label{gseq}
\frac{q \tan\frac{\theta}{2}}{1 - q}  < \gamma < \sqrt{1 - \qty[ \frac{1 - \sin{\theta} - \sqrt{2}q \sin(\frac{\pi}{4}-\theta)}{(1 - q) \cos{\theta}} ]^2}
\end{equation}
Note that $\gamma\in(0,1)$ will always exist if the lower and upper bounds of $\gamma$, denoted by $\gamma_l$ and $\gamma_u$, respectively, also lie within the range $(0,1)$. These constraints establish the upper bound of the bias parameter $q$. Given $\gamma_l  \in (0,1)$ and $q>1-q$, we obtain
\begin{equation}\label{qseq1}
    \frac{1}{2}\leq q < \frac{1}{1+\tan\frac{\theta}{2}}
\end{equation}
Similarly, the condition $\gamma_u  \in (0,1)$ implies
\begin{equation}\label{qseq2}
    \frac{1}{2} \leq q < 1-\tan\frac{\theta}{2}
\end{equation}
Since the upper bound of $q$ in Eq.~(\ref{qseq1}) is greater than that in Eq.~(\ref{qseq2}) for any $\theta\in(0,\frac{\pi}{4})$, we can, without loss of generality, restrict the permissible range of $q$ in accordance with the bound provided by Eq.~(\ref{qseq2}). Thus, there exist specific ranges of $q\in[\frac{1}{2},1-\tan\frac{\theta}{2}]$, $\gamma \in (\gamma_l,\gamma_u)$ and $\theta \in (0, \frac{\pi}{4})$ for which both Bob and Eve demonstrate nonlocality with Alice.


\subsection{Evaluation of the key rate under sequential attack}\label{Seqind}

The crux of the sequential attack strategy hinges on Eve's ability to emulate the observed CHSH violation between Alice and Bob  (without a sequential attack) by precisely adjusting the unsharpness parameters $\gamma$, biasness parameter $q$, and Alice's fabricated measurement settings $\theta$ such that $\mathcal{S}_{AB}(\alpha)=\mathcal{S}_{AB}'(q, \gamma, \theta)$. The impact of this attack is reflected in the following evaluation of the lower bound of the key rate.

This sequential measurement strategy is an individual attack, as Eve independently intercepts each system sent from Alice to Bob, applying the same method consistently. Eve measures her ancilla before classical post-processing without having access to any quantum memory. As a result, Eve, Alice, and Bob share a product distribution of classical symbols \cite{Scarani2009}.

To determine the lower bound of the key rate, denoted as $r_{S}$, we first evaluate the modified QBER as follows
\begin{equation} \label{Qseq}
\begin{aligned}
    Q^{S} &= p(a\neq b|A'_0,B_1,\rho_1) \\
    &=\frac{(1-2 Q)(1-q)}{2}\qty(1-\sqrt{1-\gamma^{2}}) + Q
    \end{aligned}
\end{equation}
where $Q$ represents the QBER of the shared correlation between Alice and Bob, as defined in Eq.~(\ref{qber}). To compare the key rate under a sequential individual attack with that of a collective attack, for a given CHSH violation and a fixed QBER, Eve must modify Alice’s initial measurement $A_0$ to $A'_0$. Note that any desired value of $Q^S$ can be obtained by choosing $A'_0$ to be $\sigma_z$ with probability $1-2Q$, and a random bit with probability $2Q$. The lower bound of the key rate is then given by \cite{Csiszar1978}
\begin{eqnarray} \label{keyS}
r_S & =& \mathcal{I}(A'_0 : B_1) - \qty[q \mathcal{I}(A'_0 : E_1)+(1-q) \mathcal{I}(A'_0:E_2)  ]  \nonumber \\
& =& 1 - H(Q^{S}) - q(1-H(Q))
\end{eqnarray}
To compare the key rate resulting from this sequential individual attack (Eq.~\ref{keyS}) with that obtained under collective attack by Eq.~(\ref{KeyRColl}), stricter bounds are required for the bias parameter $q$, Alice's measurement angle $\theta$, and the unsharp parameter $\gamma$, such that the CHSH violation (Eq.~\ref{BellABColl}) remains within the range $\mathcal{S}_{AB} \in (2,2.1213]$. 

\begin{figure}[h!]
    \includegraphics[width=\linewidth, keepaspectratio]{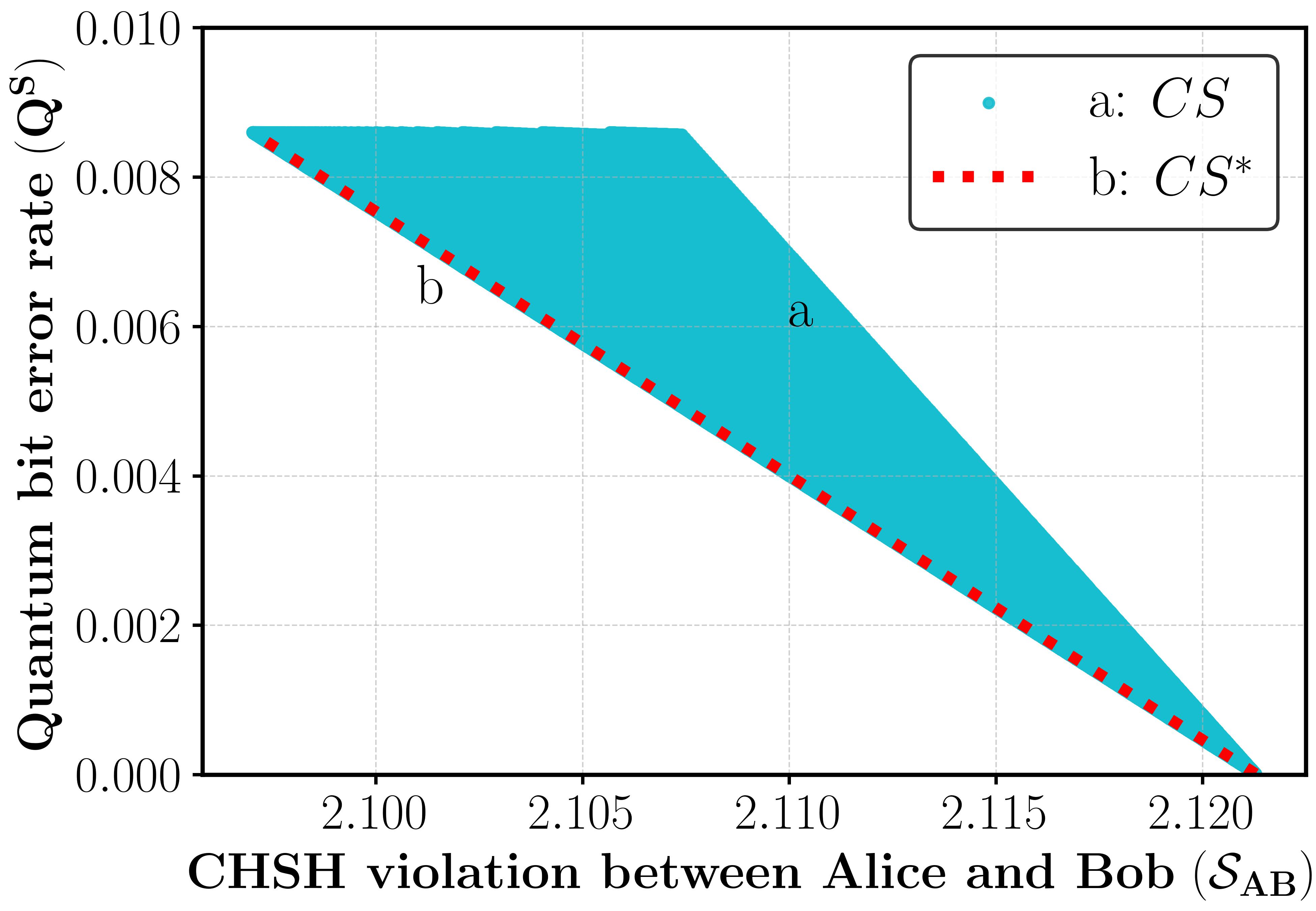}
    \caption{The quantum bit error rate ($Q^{S}$) is plotted as a function of the CHSH violation between Alice and Bob ($S_{\text{AB}}$). The cyan region highlights the range of correlations where $Q^{S}\in [0,0.0086]$ and $S_{AB}'\in [2.0965, 2.1213]$. The red dashed line represents the target correlation specified by Eqs.~(\ref{BellABColl}) and (\ref{qber}).}
     \label{qvss}
\end{figure}

\begin{figure}[h!]
\includegraphics[width=\linewidth]{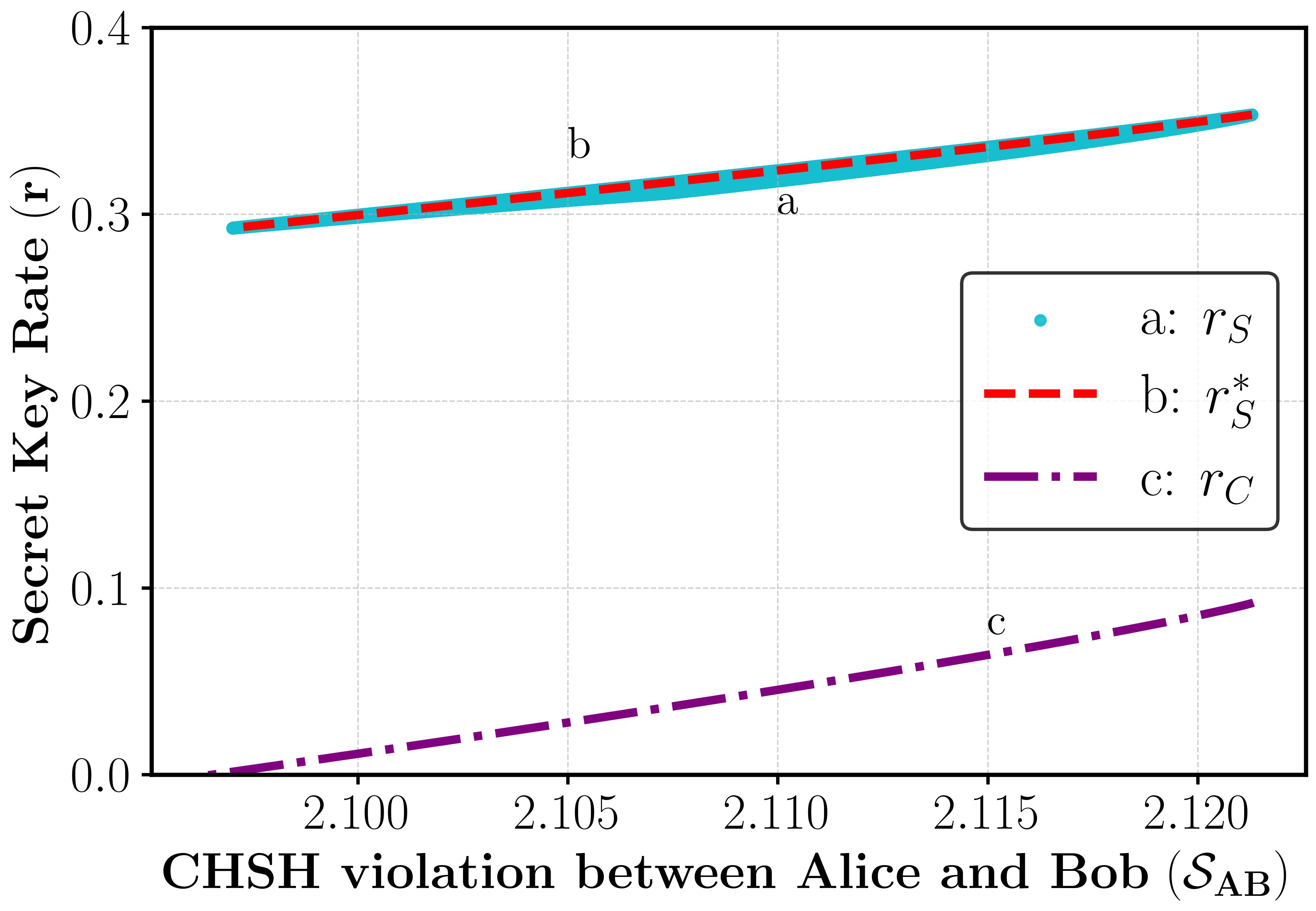}
\caption{Key rate ($r$) as a function of the CHSH violation between Alice and Bob ($S_{\text{AB}}$). Region `a' (cyan shaded) shows the key rate under a sequential attack ($r_S$) for the set of correlations highlighted in cyan in Fig.~\ref{qvss}. Curve `b' (red dotted) represents the optimal sequential attack key rate ($r_S^*$), while curve `c' (purple dash-dotted) shows the key rate under a collective attack ($r_C$) corresponding to the red dotted correlation line in Fig.~\ref{qvss}.}

\label{PlotSeq}
\end{figure}

For this purpose, Eve first manipulates $A_1$ and $A_2$ by choosing the following value of $\theta$ 
\begin{equation}
    \theta^{*}= \frac{\pi}{2} - \tan^{-1} \qty(\frac{q}{1-q}+\sqrt{1-\gamma^2})
\end{equation}
for which the CHSH value between Alice and Bob reaches its maximum, denoted as $\tilde{\mathcal{S}}^{*}_{AB}$, corresponding to the state given in Eq.~(\ref{state2}) (see Appx.~\ref{AppA} for a detailed evaluation).
Using the bounds provided in the paragraph below Eq.~(\ref{KeyRColl}), and setting $\tilde{\mathcal{S}}^{*}_{AB}=\mathcal{S}_{AB} \in [2.0965, 2.1213]$ leads to the following constraints on $\gamma$ (see Appx.~\ref{AppA}) 
\begin{equation}
\begin{cases}
\gamma_{nl}  < \gamma < \gamma_{nu}, & \text{if } 0.5< q \leq 0.6464 \\
0  < \gamma < \gamma_{nu}, & \text{if } 0.6464< q < 0.6856
\end{cases}
\end{equation}
where the expressions for $\gamma_{nl}$ and $\gamma_{nu}$ are provided in Eq.~(\ref{gammaB1}) of Appx.~\ref{AppA}. From Eq.~(\ref{keyS}), it is evident that Eve's information increases with $q$; hence, the second bound offers a more advantageous regime for Eve’s attack strategy. In this region the variation of QBER with CHSH value is illustrated in Fig.~\ref{qvss}. The red dashed line `a' represents the target correlation specified by Eqs.~(\ref{BellABColl}) and (\ref{qber}). The corresponding key rate as a function of CHSH value is presented in Fig.~\ref{PlotSeq}. We find that the key rate under the sequential individual attack exceeds that of the collective attack.

This result is expected because, in comparing individual and collective attacks, the collective attack is more powerful due to Eve’s access to quantum memory, allowing her to store her ancilla and perform a collective measurement later to extract more information \cite{Scarani2009}. In the following section, we will discuss the cumulative effect of both sequential and collective attacks.

\section{DI-QKD under the cumulative effect of sequential and collective attacks} \label{sacacom}

In a standard collective attack scenario, Eve controls the preparation device, sending Alice and Bob a mixed two-qubit entangled state while retaining the purification qubit as an ancilla. Over multiple rounds, if Eve has access to a quantum memory, she can store these ancilla states and later perform collective measurements to extract information about the shared statistics between Alice and Bob.

In the scenario considered here, however, Alice prepares the bipartite state, thereby removing Eve's control over the preparation device. Instead of retaining the purification as an ancilla, Eve intercepts the qubit sent from Alice to Bob, performs measurements following the strategy outlined in Sec.~\ref{sas}, and prepares an ancilla state based on the measurement outcomes. With access to quantum memory, Eve can store these ancilla states and subsequently perform collective measurements. This hybrid strategy combines elements of both sequential and collective attacks, requiring the key rate to be evaluated under their combined influence.

 The impact of a collective attack~\cite{Scarani2009} is characterised by the extractable information, \(\chi(B_1 : E)\), as defined in Eq.~(\ref{key}). In contrast, the effect of a sequential attack is reflected through the QBER, which depends on Eve's control parameters \(q\) and \(\gamma\), as given in Eq.~(\ref{Qseq}). To determine the optimal collective attack, one must identify the condition when \(\chi(B_1 : E) = \chi(\mathcal{S}_{AB})\), as specified in Appx.~\ref{AppB}. In this context, the lower bound of the key rate is expressed as
\begin{equation}
    r_{CS} = 1 - H(Q^S) - \chi(B_1:E)
    \label{KeyRSeqColl}
\end{equation}
where $\chi(B_1:E)$ is given by Eq.~(\ref{chiB1E}) in Appx.~\ref{AppB}. Note that, in this sequential collective attack, without having the initial state control, Eve can still hold a purification of the state as her ancilla system, as reflected in the evaluation of $\chi(B_1:E)$ in Appx.~\ref{AppB}.

When comparing the two collective attack strategies, the conventional optimal collective attack~\cite{Acin2007} and the sequential collective attack, we find that the key rate, as a function of Bell violation, becomes identical when both the Bell violation and QBER match the values given by Eqs.~(\ref{BellABColl}) and ~(\ref{qber}), respectively. This agreement is anticipated, as Eq.~(\ref{KeyRColl}) represents the ultimate lower bound for the key rate under the most general attack strategies \cite{Friedman2018}; hence, no attack can reduce the key rate below this threshold.

A key distinction lies in the assumptions concerning Eve’s capabilities. In the traditional optimal collective attack, Eve must have initial control over the preparation device. In contrast, the sequential collective attack considered here assumes no such initial control. Instead, Eve exploits a carefully tailored sequential measurement strategy to influence the preparation process indirectly. Depending on her resources and objectives, she may opt for a sequential individual attack, as discussed earlier in Sec.~\ref{Seqind}, or, if equipped with a quantum memory, she can implement an optimal sequential collective attack.

\begin{figure}[h!]
\centering\includegraphics[width=\linewidth]{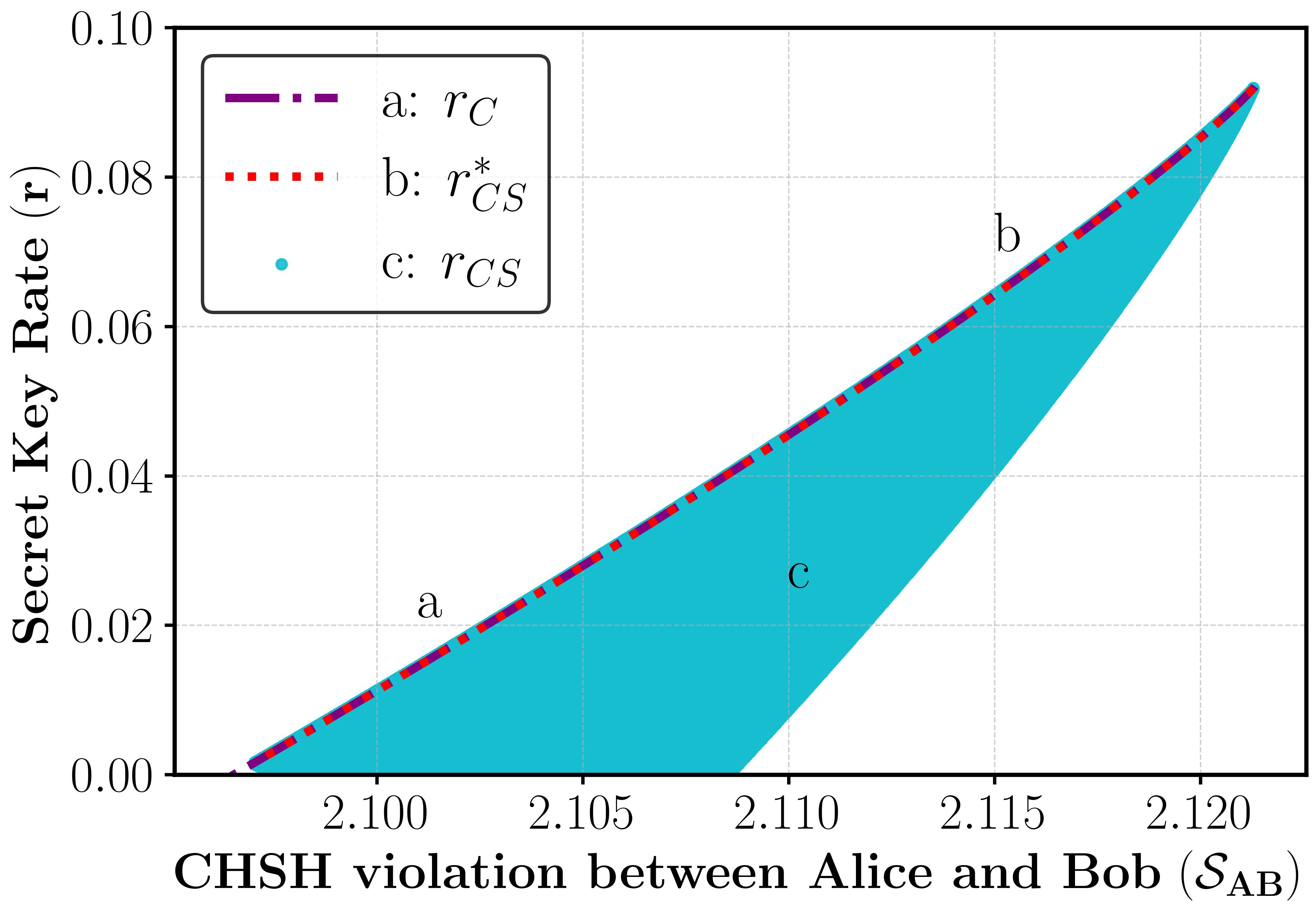}
\caption{Secret key rate ($r$) as a function of the CHSH violation between Alice and Bob ($S_{\text{AB}}$). Curve ‘a’ (purple dash-dotted) represents the key rate under a collective attack ($r_{C}$), while curve ‘b’ (red dotted) corresponds to the optimal sequential collective attack ($r_{CS}$). Region ‘c’ (cyan shaded) highlights the set of correlations that yield a positive secret key rate, constrained by a quantum bit error rate $Q^{S} \in [0, 0.0086]$ and CHSH values $S'_{\text{AB}} \in [2.0965, 2.1213]$, as defined by the red dotted line in Fig.~\ref{qvss}.}
\label{graph0}
\end{figure}

In Fig.~\ref{graph0}, curve `a' shows the key rate as a function of CHSH value $(\mathcal{S}_{\text{AB}})$ under the optimal collective attack, based on the correlation defined by Eqs.~(\ref{BellABColl}) and (\ref{qber}). Curve `b' shows the key rate under the optimal sequential collective attack, derived from the same correlation. Additionally, in region 'c' in Fig.~\ref{graph0} represents the set of key rates corresponding to all possible correlations for which the CHSH violation satisfies $\mathcal{S}^{'}_{AB}\in[2.0965, 2.1213]$ and the QBER satisfies $Q^S \in [0, 0.0086]$, pertaining to the cyan region in Fig.~\ref{qvss}.

In our analysis, the CHSH violation is notably low, which raises questions regarding its relevance in the context of DI-QKD. Nevertheless, recent studies demonstrate that DI-QKD remains feasible even with minimal non-locality \cite{Wooltorton2024}. Additionally, experimental findings confirm that the CHSH violation threshold aligns with the conditions considered in our case \cite{Xiao2024ExptSharing}.


\section{Salient Features and Outlook}\label{Salient}

 With the advancement of loophole-free Bell experiments, DI-QKD is approaching practical implementation. While existing DI-QKD security proofs have accounted for the possibility of Eve hiding eavesdropping attempts in the noise levels tolerated by the protocol, some physical insight could be gained by studying explicit examples of such eavesdropping attempts.

In most DI-QKD scenarios, Eve is typically assumed to have control over both the entangled source and measurement devices. In our work, we consider a practical setting common in experimental or commercial DI-QKD, where the entangled source resides at Alice’s node. To maintain the DI paradigm, we assume Eve can tamper with Alice’s measurement device during manufacturing but lacks direct control over the source.

The central question we examine is whether Eve can reproduce the behaviour of known powerful attacks~\cite{Scarani2009,Acin2007} even when she does not control the source. This does not challenge the principles of device independence; rather, it isolates a realistic situation in which Eve’s influence is limited to the travelling system. Our analysis focuses on a concrete scenario where Eve’s only operation is a sequential intervention on the transmitted qubit. This assumption reflects practical constraints in certain DI-QKD implementations and provides a well-defined setting in which to analyse how such restricted adversarial actions manifest at the level of observable statistics, without implying any additional security weakness.

Our analysis considers a scenario where Eve intercepts the qubit travelling from Alice to Bob, performs an unsharp measurement, and generates an ancilla conditioned on her outcome. She may store the ancilla states and later apply collective measurements. This attack does not affect the security of DI-QKD, as it is covered by existing security proofs, but it gives a concrete, physically implementable operation Eve can perform within that scope.

In this work, we considered the asymptotic key rate; however, the effectiveness of the sequential interaction depends on Eve’s bias parameter $q$, which would in practice be limited by experimental constraints. Finite-size security for DI-QKD in the presence of general adversarial actions, including sequential operations, is already covered by entropy accumulation techniques~\cite{Dupuis2020EAT, Friedman2018}, so our analysis should be viewed as providing a concrete illustration rather than extending these results. A natural direction for future work is to study how sequential unsharp-measurement strategies extend to DI-QKD protocols that use more than two inputs per party. Increasing the number of inputs typically strengthens the corresponding Bell inequality and may enhance key rates, but it also requires adapting the sequential measurement so that each observer’s action remains physically implementable while still preserving the required nonlocal correlations. Exploring how to construct such practically achievable sequential strategies for higher-input nonlocality-based QKD protocols could provide further insight into the operational structure of sequential measurements, which seems to be a promising direction for future analysis.


\section{Acknowledgements} 
S.S. acknowledges support by the National Natural Science Fund of China (Grant No. G0512250610191). S. G. acknowledges OIST, Japan for the postdoctoral fellowship.


%


\appendix
\onecolumngrid

\section{Relevant Upper and lower bounds of parameters $\gamma$, $q$ and $\theta$}\label{AppA}

To determine the range of the parameters $\gamma$, $q$, and $\theta$ that enable Eve to reproduce the observed correlation, she first selects the angle $\theta$ in such a way that Alice and Bob obtain the maximum violation of CHSH, corresponding to the state given in Eq.~(\ref{state2}).

In the sequential attack, lets consider $\gamma_1 = 1$ and $\gamma_2 = \gamma$, which transforms the state in Eqn.~(\ref{state2}) into
\begin{equation}\label{StateNew}
    \rho_1 = \Lambda_{\phi^+}\ket{\phi^+}\bra{\phi^+} + \Lambda_{\phi^-}\ket{\phi^-}\bra{\phi^-} + \Lambda_{\psi^+}\ket{\psi^+}\bra{\psi^+},
\end{equation}

where the coefficients are given by

\[
\Lambda_{\phi^+} = \frac{1}{2}\qty[1+(1-q)\sqrt{1-\gamma^2}], \quad \Lambda_{\phi^-} = \frac{q}{2}, \quad \Lambda_{\psi^+} = \frac{1-q}{2}\qty(1-\sqrt{1-\gamma^2}).
\]

Using the Horodecki criterion~\cite{horodecki1995violating}, we find that the optimal CHSH value for the state of Eq.~(\ref{StateNew}) is given by
\begin{equation}
\begin{aligned}\label{Sopt}
    \tilde{\mathcal{S}}_{AB}^{*} &= 2\sqrt{(\Lambda_{\phi^-}-\Lambda_{\phi^+}-\Lambda_{\psi^+})^2+(\Lambda_{\phi^-}+\Lambda_{\phi^+}-\Lambda_{\psi^+})^2} \\
    &= 2 \sqrt{2 - \gamma^2 - 2q \left(2 - \gamma^2 - \sqrt{1 - \gamma^2}\right) + q^2 \left(3 - \gamma^2 - 2 \sqrt{1 - \gamma^2}\right)}.
\end{aligned}
\end{equation}
By comparing Eq.~(\ref{BellABSeq}) with Eq.~(\ref{Sopt}), we can express the angle $\theta$ as
\begin{equation}\label{thetaopt}
\theta^{*} = \frac{\pi}{2}-\tan^{-1}\left(\frac{q}{1-q}+\sqrt{1-\gamma^2}\right).    
\end{equation}
By constraining the CHSH violations within the interval $[2.0965, 2.1213]$, as prescribed by Eq. (\ref{Sopt}), we get two different regions in parameter space $\gamma$ and $q$ ,

\textit{Region 1:}
When $0.5 < q \leq 0.6464$, and 
$\gamma_{nl}  < \gamma < \gamma_{nu}$. Where $\gamma_{nl}$ and $\gamma_{nu}$ is given by, 
\begin{equation}
\label{gammaB1}
    \gamma_{nl} = \sqrt{1 - \qty[ \frac{0.5}{1 - q}\sqrt{0.4999 + 8 q - 4 q^2}- \frac{q}{1 - q} ]^2}
     \ \ \text{and} \ \ 
    \gamma_{nu} = \sqrt{1 - \qty[ \frac{0.25}{1 - q} \sqrt{
    1.5813 + 32 q - 16 q^2} -\frac{q}{1 - q}]^2}
\end{equation}

\textit{Region 2:}
When $0.6464 < q < 0.6856$, and $0<\gamma<\gamma_{nu}$. $\gamma_{nu}$ is same as Eq.~(\ref{gammaB1}).

\section{Eve's information in collective attack}\label{AppB}
We know that, $\chi(B_1 : E)= S(\rho_E) - \frac{1}{2}\sum_{b_1 = \pm 1}S(\rho_{E|b_1})$. It is shown in Ref.~\cite{Pironio2009}, for bell diagonal state $\chi(B_1 : E)$ is given by,
\begin{equation}
    \chi(B_1 : E) = - \sum_{i = \Lambda_{\phi^+}, \Lambda_{\phi^-}, \Lambda_{\psi^+}} \Lambda_{i} \log_{2} \Lambda_{i} - \frac{1}{2}\left(S(\rho_{E|b_{1}=+1})+S(\rho_{E|b_{1}=-1})\right)
\end{equation}
A similar argument as in Ref.~\cite{Pironio2009} shows that $S(\rho_{E|b_1=+1}) = S(\rho_{E|b_1=-1}) = H(\Lambda_+)$, where $\Lambda_+ $, the largest eigenvalue of $\rho_{E|b_1}$, is given by:
\begin{align*}
    \Lambda_{+} = \frac{1}{2}\left(1 + \sqrt{\Lambda_{\phi^+}^2 + (\Lambda_{\phi^-} - \Lambda_{\psi^+})^2 + 2 \Lambda_{\phi^+} (\Lambda_{\phi^-} - \Lambda_{\psi^+})}\right)
\end{align*}
Eve's information in a collective attack is given by
\begin{equation}
\label{chiB1E}
    \chi(B_1 : E) = - \sum_{i = \Lambda_{\phi^+}, \Lambda_{\phi^-}, \Lambda_{\psi^+}} \Lambda_{i} \log_{2} \Lambda_{i} - H(\Lambda_+),
\end{equation}
As shown in Ref.~\cite{Pironio2009}, to achieve an optimal collective attack, one must maximize Eq.~(\ref{chiB1E}). It was proven that this quantity is maximized when either \(\Lambda_{\phi^+} = \Lambda_{\phi^-} = 0\) or \(\Lambda_{\psi^+} = \Lambda_{\psi^-} = 0\).

In the context of our sequential collective attack scenario, \(\Lambda_{\psi^-} = 0\) is already guaranteed by Eq.~(\ref{StateNew}). Therefore, optimality in this case requires \(\Lambda_{\psi^+} \rightarrow 0\), which implies \(\gamma \rightarrow 0\).

\end{document}